\documentclass{jaa}
\usepackage{graphicx}
\begin{document}
\title{Observing Compact Stars with AstroSat}
\author{Dipankar Bhattacharya\textsuperscript{1,*}}
\affilOne{\textsuperscript{1}Inter-University Centre for Astronomy and Astrophysics, Pune 411007, India}
\twocolumn[{
\maketitle
\corres{dipankar@iucaa.in}
\msinfo{1 July 2017}{6 July 2017}{6 July 2017}
\begin{abstract}
This article presents a brief description of India's AstroSat mission which is a powerful space based observatory for compact star research.  An account is given of observational constraints and spectral and timing capabilities as realised post-launch. Some preliminary results of observations of the Crab pulsar and an X-ray binary system GX~301-2 are presented to illustrate some of the capabilities of the mission. 
\end{abstract}
\keywords{Compact Stars---Multi-wavelength observations---AstroSat}
}]

\doinum{}
\artcitid{\#\#\#\#}
\volnum{}
\year{2017}
\pgrange{}
\setcounter{page}{1}
\lp{8}

\section{Introduction}
AstroSat, India's first space-borne observatory (Singh {\em {\em et al}} \nocite{singh+2014} 2014), was launched on 28 September 2015.  The mission was conceived during the period 1996-2000 and built over the next decade and a half. Several key aspects of the mission design make AstroSat a powerful instrument for the study of compact stars.  A great deal of knowledge about compact stars is gathered from the study of accreting systems, where a donor star supplies matter to the compact star, and the accreted matter generates the observed radiation.  As the matter approaches the compact object, its temperature gradually increases; the inner parts of the flow thus glow in high energy X-rays while the outer parts produce ultraviolet emission.  AstroSat carries four co-aligned science payloads that provide simultaneous coverage of this wide energy band, from near and far UV to hard X-rays up to $\sim 100$ keV.  Radiation from the accretion flow is strongly variable; the nature and the time scales of such variability carry a wealth of information about not only the flow but also the compact star itself (see, e.g. \nocite{hk:89} Hasinger and van der Klis 1989, \nocite{bm:16} Belloni and Motta 2016).  AstroSat provides an unprecedented capability to study such rapid variability simultaneously at all available wavebands. All science payloads on AstroSat record individual photon events with sub-second timing resolution, at medium to hard X-rays the time resolution being as short as 10--20 $\mu$s.  As photon energies are also registered for each event,  multi-wavelength, spectrally resolved time variability can be studied in unmatched detail with this mission.  Broadband spectroscopy can be carried out and spectral variability can be studied with high sensitivity.  These capabilities were chosen to serve the long-standing interest of the Indian community in the area of compact stars, including those pursued over many years by the research group led by G. Srinivasan: for example the high energy emission from pulsars (Srinivasan \nocite{srini:1990} 1990), the formation and evolution of pulsars (Srinivasan  \nocite{srini:1989} 1989, Srinivasan {\em et al} \nocite{sdb:1984} 1984), the evolution of the magnetic fields of neutron stars (Srinivasan \nocite{srini:1991} 1991, Srinivasan {\em et al} \nocite{srini+1990} 1990), the structure of neutron stars (Srinivasan \nocite{srini:2002} 2002) and particularly the accretion-driven evolution of neutron stars (Srinivasan \nocite{srini:2010} 2010).  In this paper I briefly outline the characteristics of AstroSat instruments, discuss the post-launch performance and observing constraints and present some preliminary results to highlight the mission capabilities for compact object research.

\section{AstroSat science instruments}
AstroSat carries five science instruments on board.  Among them the following four are co-aligned and view a given target concurrently.

\paragraph{(i) LAXPC:}  The Large Area X-ray Proportional Counters (LAXPC) consist of three proportional counter units, with a detection volume of 100~cm $\times$ 36~cm $\times$ 15~cm each, filled with Xenon-Methane mixture in two units and Xenon-Argon-Methane mixture in one unit, at a pressure of  about 2 atmospheres.  Each detector is fitted with a $\sim$45~cm high collimator stack providing window support and $\sim 1^{\circ}\times 1^{\circ}$ restriction of the field of view.   The combined effective area of the three units is $\sim 6000$~cm$^2$ in 5$-$20~keV band, dropping to $\sim 5000$~cm$^2$ at 50~keV and $\sim 1000$~cm$^2$ at 100~keV. The instrument records photon events with a time resolution of 10~$\mu$s.  The spectral resolution is $\sim 12$\% above 20~keV and worsens to about 24\% at 5 keV.  Detailed description of the LAXPC instrument and its performance can be found in \nocite{antia+2017} Antia {\em et al} (2017). 

\paragraph{(ii) CZTI:} The Cadmium Zinc Telluride Imager (CZTI) instrument uses a 5-mm thick solid-state pixellated Cadmium Zinc Telluride hard X-ray detector array with ~$\sim 48$~cm high collimators defining a $4.6^{\circ} \times 4.6^{\circ}$ field of view in the 20-100~keV range, and a coded aperture mask on top  providing imaging capability.  The instrument is divided into four independent quadrants.   The total geometric area of the detector is 976~cm$^2$, distributed over 16384 pixels.  After launch, about 15\% of the pixels showed excessive electronic noise and have been disabled.  About 25\% of the remaining pixels were found to have poor spectroscopic response.  Given the $\sim 50$\% transparency of the coded mask, the total effective area at normal incidence is about 420~cm$^2$ in all active pixels, and about 315~cm$^2$ in spectroscopically good pixels.  In the latter, the typical energy resolution is $\sim 6$\% and the low energy threshold is between 20--30~keV.  The detectors are sensitive to energies up to 250~keV, but the mask and the collimator become progressively transparent above 100 keV.  CZTI records detected events with a time resolution of 20~$\mu$s.  A description of the CZTI instrument may be found in \nocite{bhalerao+2017} Bhalerao {\em et al} (2017).

\paragraph{(iii) SXT:} The Soft X-ray Telescope (SXT) employs a gold-coated foil mirror grazing incidence reflecting optics consisting of 40 concentric shells, with a focal length of 2~m.  An X-ray CCD camera located at the focus provides a resolution of 600$\times$600 pixels, each of $\sim 4$~arcsec square.  The on-axis point spread function has a full width at half maximum of 100~arcsec and a 50\% encircled energy diameter of 11 arcmin.  The operational energy range of SXT is 0.3 to 8~keV, with a typical energy resolution of $\sim 150$~eV.  Two modes of readout are available to the user, a Photon Counting (PC) mode of the full CCD frame with a time resolution of 2.4~s, and a Fast Window (FW) mode that reads only the central 150$\times$150 pixels with a time resolution of 0.278~s, which is the highest time resolution available with the SXT.  The peak effective area is $\sim 120$~cm$^2$ in 0.8--2~keV range; over 2.5--5~keV the area is $\sim 60$~cm$^2$ and drops gradually to $\sim 7$~cm$^2$ at 8~keV.  Details of the SXT and its operations are described in Singh {\em et al} \nocite{singh+2017} (2017).

\paragraph{(iv) UVIT:} The Ultraviolet Imaging Telescope (UVIT) is a combination of two Ritchey-Chretien reflecting telescopes, each with a primary mirror of 38~cm diameter.  One telescope has a Far Ultraviolet (FUV) detector unit at the focus while the other contains a dichroic that splits the beam into Near Ultraviolet (NUV) and Visible (VIS) channels which have their own independent detectors.  Thus UVIT can observe simultaneously in all these three bands.  The VIS channel is primarily used for image tracking required to provide drift corrections. Photometry is not well calibrated in this channel.  FUV and NUV are the main science bands. Several filters, as well as a grating, are available in either of them.  The detector in each channel consists of a microchannel plate followed by a 512$\times$512 pixel CMOS photon counting detector, providing a field of view of $\sim 28$~arcmin square. Coordinates of the events recorded in the full frame are read off and time-tagged 29 times a second.  If faster readout 
is required, one of several smaller window selections can be used, the smallest being a 100$\times$100 pixel window which can be read and time-tagged at 1.6~ms intervals. The drift-corrected angular resolution of UVIT in the FUV and NUV bands are approximately 1.4~arcsec. Tandon {\em et al} \nocite{tandon+2017}  (2017) provide an overview of the UVIT, its operations and performance.

The fifth science instrument, the Scanning Sky Monitor (SSM) is mounted on a rotating platform. The axis of rotation is oriented orthogonal to the pointing direction of the co-aligned instruments.  This instrument consists of three position sensitive proportional counter detectors with one-dimensional coded masks, akin to the All Sky Monitor aboard NASA's Rossi X-ray Timing Explorer mission (Levine {\em et al} \nocite{levine+1996} 1996). Each detector has a geometric area of $\sim 60$~cm$^2$, and an operating energy range of 2--10~keV.  In one rotation of the platform the field of view of these cameras sweep through a full hemisphere of the sky.  There is also provision to stop the rotation and obtain long exposures on a desired target.  The SSM is described in detail by Ramadevi {\em et al} \nocite{ramadevi+2017} (2017). 

In the satellite coordinates, the common pointing axis of the co-aligned instruments is referred to as the "Roll" axis.  The axis of rotation of the SSM platform is the "Yaw" axis, and the third satellite axis orthogonal to both Roll and Yaw is called the "Pitch" axis.  Although the Roll axis is nominally the common view axis of the co-aligned instruments, there are inter-instrument mounting offsets amounting to several arc minutes.  For each observation a "primary instrument" is therefore defined based on the science requirement, and the view axis of the primary instrument is pointed to the target, not the Roll axis itself.  If a small window observation is chosen for SXT or UVIT, then one must ensure that the corresponding instrument has been designated the primary instrument, otherwise the desired target may fall outside the observing window. 

\section{Observational constraints}

Observation of a desired target by AstroSat is subject to several operational constraints encountered by the mission.  These constraints determine the scheduling of the observations, and in some cases result in strong restrictions on visibility.  The primary constraints are as follows.

\paragraph{1. Sun Angle:} Most of the science payloads on board AstroSat need to avoid direct view of the Sun as well as scattered sunlight. Instruments most strongly affected by the Sun are the UVIT and the SXT, and based on their requirements an overall mission constraint of Sun angle larger than 45$^{\circ}$ from the Roll axis has been defined.  In addition, the SSM field of view must avoid the Sun as well, so the satellite is always rotated to keep the Sun in the negative Yaw hemisphere.  Further, to ensure the best thermal control and stability, the satellite usually operates with the Sun kept at an angle between 65$^{\circ}$ and 150$^{\circ}$ from the positive Roll direction.
 
\paragraph{2. Moon Angle:} An overall constraint of avoidance of the moon within 15$^{\circ}$ of the Roll axis is implemented in AstroSat observations.
  
\paragraph{3. Earth Limb Angle:} An avoidance zone of 12$^{\circ}$ around the bright limb of the Earth is defined for AstroSat operations.

\paragraph{4. Ram Angle:} In order to avoid damage to the UVIT and SXT mirrors arising from direct bombardment of atomic oxygen as the satellite moves through the local atmosphere, it has been stipulated that the angle between the velocity vector of the satellite and the view axis of these instruments must never fall below 12$^{\circ}$.  This does not place any constraint on observations of  targets more than 12$^{\circ}$ away from the satellite's orbital plane.  For others, the constraint can in principle be avoided by manoeuvring away from the target for a small section of each orbit and then returning to it.  However, based on the performance of the reaction wheels on board AstroSat the Mission control has determined that such frequent manoeuvres are not desirable.  Targets falling within the Ram angle avoidance zone can therefore not be observed by AstroSat.  The orbital inclination of AstroSat is 6$^{\circ}$ with respect to the Earth's equator, and this orbit precesses with a period of $\sim 50$~days.  As a result, targets with declination $\delta$ within $\pm 18^{\circ}$ are affected by the Ram angle constraint.  Those with $6^{\circ} < |\delta| < 18^{\circ}$ become visible for some part of the precession cycle, but targets with $|\delta| < 6^{\circ}$ are permanently outside AstroSat's view.

\paragraph{5. Brightness of target and field:}  Extremely bright targets can cause damage to the detectors of some of the AstroSat's science instruments, and should therefore be avoided.  Scorpius X-1, a bright X-ray binary, is thus excluded from the direct view list of the LAXPC.  The brightness constraints are most severe for the UVIT instrument.  The combination of filters in different channels need to be carefully chosen to keep the count rates within acceptable limits.  If no appropriate filter setting can be found to satisfy this then the target cannot be observed by UVIT.  When excessive brightness is detected during an ongoing observation, the UVIT generates a BOD (Bright Object Detect) signal and shuts off automatically.  Restoring normal operations after a BOD event involves a time-consuming manual operation.

\begin{figure}
\includegraphics[angle=270,width=\columnwidth]{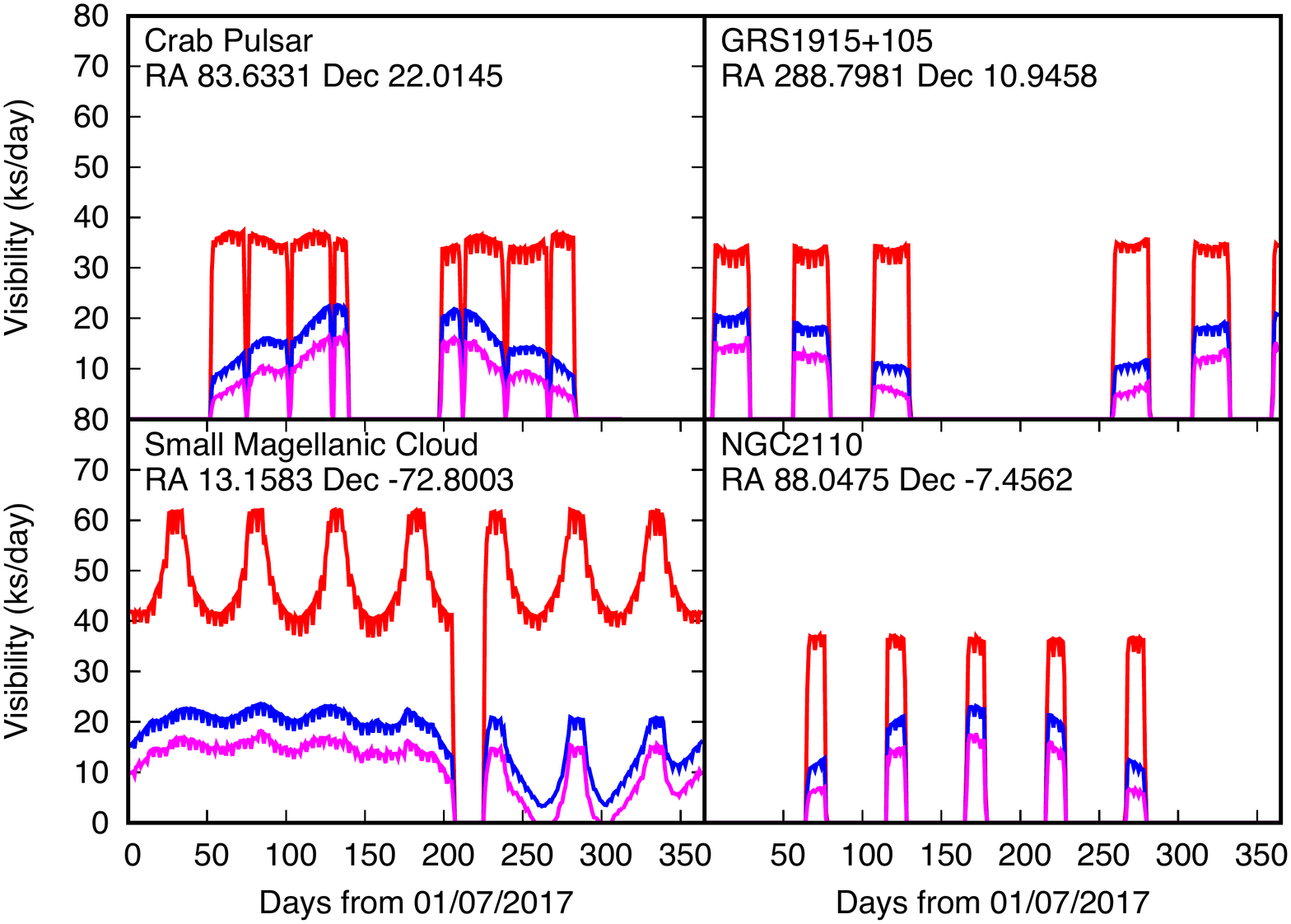}
\caption{Visibility to AstroSat of sources in different parts of the sky.  Visibility in kiloseconds during each day over a one-year period is shown.  Three curves are shown in each panel, from upper to lower, corresponding to visibility durations in LAXPC or CZTI (red), SXT (blue) and UVIT (magenta) instruments respectively, The two sources in the right hand panels are Ram angle constrained, while those on the left are not.  SMC, which is at a high negative declination, is visible for a large part of the year but for a small interval excluded by solar proximity. For the Crab the large gaps in visibility are caused by Sun angle constraint and small gaps at one-month intervals are due to the Moon.}
\label{fig:visibility}
\end{figure}

AstroSat orbits the Earth about 14 times a day.  The observing schedule normally consists of a series of long duration pointings lasting multiple orbits and may at times be punctuated by short observations lasting a single orbit.  The number of manoeuvres is limited to a few ($<5$) per day.  As a result, the duty cycle of observations is limited by target occultation by the Earth and passage through the South Atlantic Anomaly.  During the latter, all the detectors are kept off to avoid damage by accelerated charged particles.  In addition, the UVIT and the SXT are able to operate only in the night side of the orbit, further reducing the duty cycle.  The available observing time for UVIT in an orbit is even smaller than that of the SXT because for it both the start-up and the shut-down sequence must be carried out while the satellite is still in the orbital night.   On an average, the duty cycle of observation of the LAXPC and CZTI payloads is about 45\%, that of the SXT $\sim 25$\% and of the UVIT $\sim 15$\%. Figure~\ref{fig:visibility} shows examples of source visibility in different AstroSat instruments over a one year period.   Constraints such as these need to be kept in mind while proposing for observations of any target with AstroSat.

\section{Observations and performance}
As a group, compact objects enjoy the maximum usage of AstroSat time.  During the first one and a half years of operation, nearly half of the mission time has been devoted to observations of white dwarfs, neutron stars, and stellar mass black holes; another $\sim 15$\% has been spent on active galaxies harbouring supermassive black holes.  This trend is likely to continue through the life of the mission.  The main objectives of most of the compact object observations have been X-ray timing and broad band spectroscopy, including the study of cyclotron lines.  While the analysis of much of the data is still in process, some of the first results have been announced.  Energy-resolved timing features of Black Hole Binaries GRS~1915+105 and Cygnus~ X-1 over 3--80 keV band have been studied in detail using the LAXPC (Yadav {\em et al} \nocite{yadav+2016} 2016, Misra {\em et al} \nocite{misra+2017} 2017).  In LAXPC observations of the neutron star X-ray binary 4U~1728-34, kilohertz quasi-periodic oscillations with drifting frequency, X-ray burst and burst oscillations have been detected (Chauhan {\em et al} \nocite{chauhan+2017} 2017).  The CZT Imager instrument has carried out phase resolved polarimetry of the Crab Pulsar in 100--380 keV energy band (Vadawale {\em et al} \nocite{vadawale+2017} 2017), shedding new light on the production of high energy emission in fast spinning pulsars -- with possible implications on the generation of gamma rays from millisecond pulsars (Srinivasan \nocite{srini:1990} 1990).  CZTI also routinely observes Gamma Ray Bursts (Rao {\em et al} \nocite{rao+2016} 2016, Basak {\em et al} 2017) and has detected more than 100 till date.  The brightest of them are being examined for hard X-ray polarisation (Chattopadhyay {\em et al} \nocite{chat+2017} 2017).

One of the key technical elements in the study of multi-wavelength rapid variability with AstroSat is the accuracy and stability of the clocks of the different instruments and the relative alignment of their time stamps.  An UTC time reference on AstroSat is provided by a Spacecraft Positioning System (SPS) which makes use of signals from GPS satellites for its operation.  Once in 16 seconds a synchronising pulse is sent to all the X-ray instruments on board AstroSat as well as the SPS and the corresponding time values of the local clocks of all these instruments are recorded along with the UTC in a Time Correlation Table (TCT).  Time stamps for the events recorded by individual instruments are derived from their local clocks. UTC time stamps are then assigned to these events via interpolation in the Time Correlation Table.  Comparison of the instrument times and the corresponding SPS times in the TCT reveals the behaviour of the local clocks.  The difference of the instrument time from the SPS time exhibits, in general, a three component behaviour: a secular, nearly linear drift, a cyclic variation with the satellite orbit and a random jitter.  The cyclic orbital component is discernible in all instruments except the LAXPC. In AstroSat data analysis, the secular and the cyclic trends are compensated for by piecewise linear interpolation between successive samples in the TCT.  Residual clock jitter after de-trending have a typical rms value of $\sim 4 \, \mu$s for LAXPC, $\sim 3 \, \mu$s for CZTI and $\sim 40 \mu$s for SXT.  The UVIT does not use the 16~s synchronising pulse; its Time Correlation Table is built from independent clock samples of each channel recorded every 1.024~s.  These too show secular and orbital variations with respect to the corresponding UTC values recorded in the TCT.  After de-trending, however, the residuals show random jumps of amplitude a few ms to about a second over time scales of minutes to hours.  While these may not significantly affect the study of aperiodic variability such as quasi-periodic oscillations, phase coherent analysis of fast pulsars with UVIT remains a challenge.

Assigning time stamps in the manner mentioned above does not preclude an overall fixed clock offset of an instrument with respect to the true absolute time.  Extensive effort has been made to determine the fixed clock offsets of LAXPC and CZTI instruments with respect to a global reference, using  observations of the Crab pulsar spread over a year and a half.  Pulse profiles generated from AstroSat instruments were compared with those from ground based radio observatories, including the Ooty Radio Telescope where daily observations of the Crab pulsar were arranged, as well as the Giant Metrewave Radio Telescope and the Jodrell Bank radio observatory.  The fixed offsets of CZTI and LAXPC clocks have thus been determined to be less than 2~ms with respect to the Jodrell Bank reference (A. Basu {\em et al} 2017, in preparation).  

If sensitive observations of low-amplitude rapid variability is one's science goal, then care must be taken to account for the effects of pointing jitter.  The effective area of every instrument is dependent on the angle of incidence, and this can introduce variability in the observed counts as the pointing axis moves around.  In AstroSat pointing stability is achieved by an attitude control servo system involving two Star Sensors and gyros.  Star sensor data are usually available only for a part of each orbit; for the rest of the time gyro data are used for attitude sensing.  While the Star sensors are in the control loop, the attitude variations are contained within $\sim 3$~arcmin of the nominal pointing direction.  When only the gyros are being used, the amplitude of the attitude variations can be a factor of two to three larger.  Attitude values sampled at 128 millisecond interval are made routinely available as auxiliary data.  More precise determination of attitude variations can be obtained from ground analysis of UVIT data;  however the UVIT is operational for only a small fraction of each orbit.

Spectral response of the AstroSat instruments have been determined from in-flight measurements.  For the LAXPC and the CZTI the analysis software dynamically generate the spectral response specific to a given observation.  The generated response matrices are estimated to have a systematic error of $\sim2$\%.  The background spectrum for LAXPC is also generated by the analysis software, and is at present thought to have a systematic uncertainty of $\sim 3$ to 5\%. The  CZTI, on the other hand, provides a simultaneous measurement of the background due to the presence of the coded mask, and routine analysis produces a background-subtracted spectrum of the source.  The response of the SXT has been well calibrated for on-axis sources and work is ongoing to refine the off-axis responses.  At count levels larger than $\sim100$~s$^{-1}$ the SXT observations begin to suffer from pile-up; appropriate corrections need to be applied for the spectral analysis of such sources.  For the UVIT, photometric calibration has been established for all the filters, for observations conducted in the Photon Counting mode.  This is applicable to the FUV and the NUV channels.  The VIS channel, however, operates in Integration Mode which does not yield reliable photometric estimates.

\section{Example Results}

\begin{figure}
\includegraphics[angle=270,width=\columnwidth]{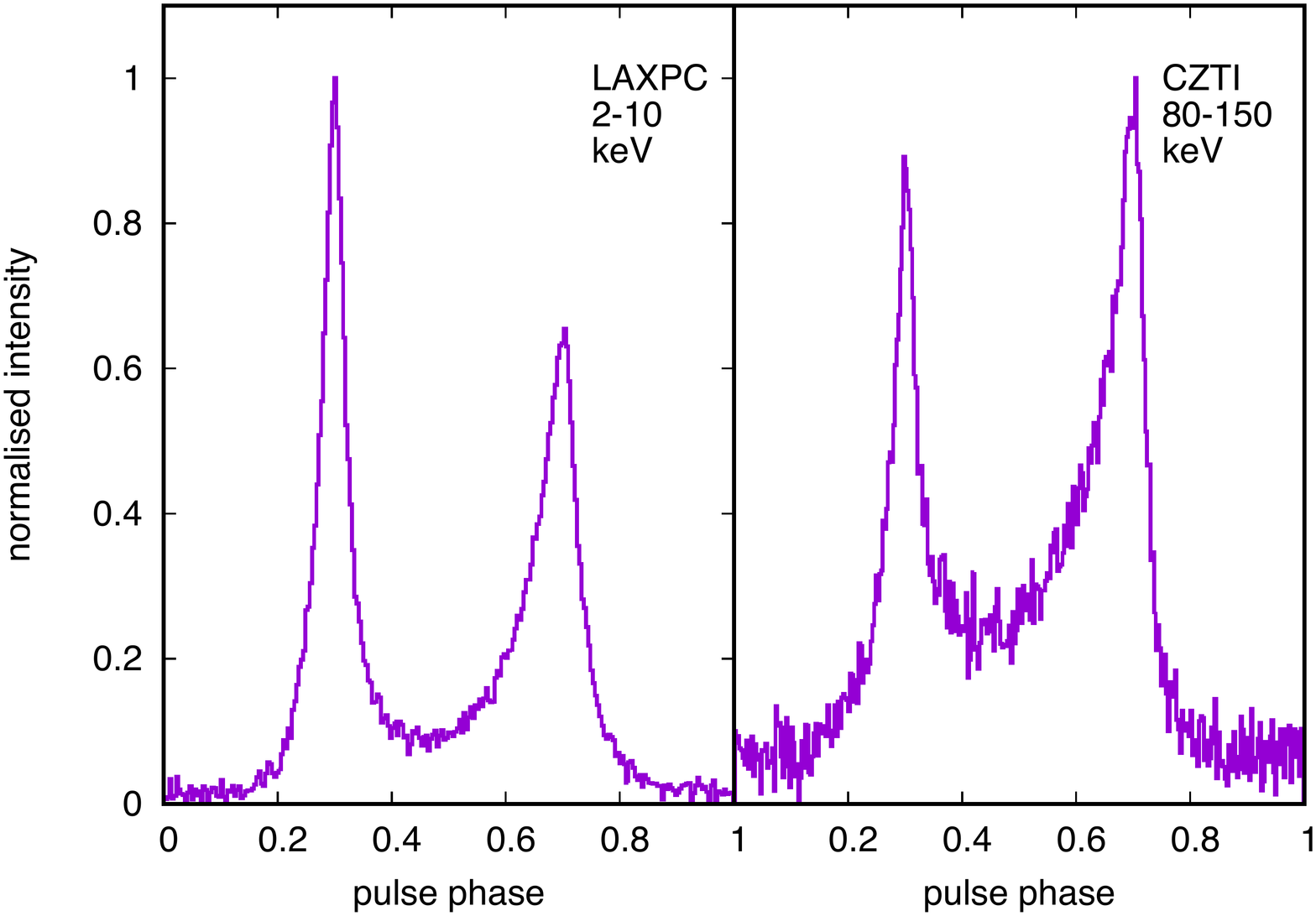}
\caption{The pulse profile of the Crab pulsar observed with one LAXPC unit (left) and the CZTI (right)  aboard AstroSat.  The integration time in the 2--10~keV LAXPC profile is $\sim 3$~ks, while that in the 80--150~keV CZTI profile is $\sim 400$~ks.  Energy dependence of the pulse profile is clearly seen: the left peak is taller at low energies while the right peak dominates at high energies. The bridge emission connecting the two peaks is also relatively stronger at higher energies.}
\label{fig:crabpsr_xray}
\end{figure}

I present here some results of the observations of the Crab pulsar and the high mass X-ray binary GX~301-2 to illustrate some of the capabilities of AstroSat relevant for compact object studies. Figure~\ref{fig:crabpsr_xray} shows the folded pulse profile of the Crab pulsar (Period: $\sim 33$~ms) using the LAXPC and the CZTI instruments.  The event time stamps were converted to Barycentric arrival times before folding, and the nearest available radio ephemeris was used for the pulsar's spin parameters.  The LAXPC profile is obtained from data collected over a single AstroSat orbit, with an effective exposure of about 3~ks.  The CZTI profile accumulates data from multiple observations spread over several months, with a total effective exposure time of $\sim 400$~ks.  The figure displays profiles in two widely separated energy bands, and the energy dependence of the shape of the profile is clearly seen.  This result demonstrates the stability of event time stamps, the accuracy of barycentric corrections that involve the knowledge of the satellite's orbit and the capability of long-term phase-connected analysis.  

\begin{figure}
\includegraphics[width=\columnwidth]{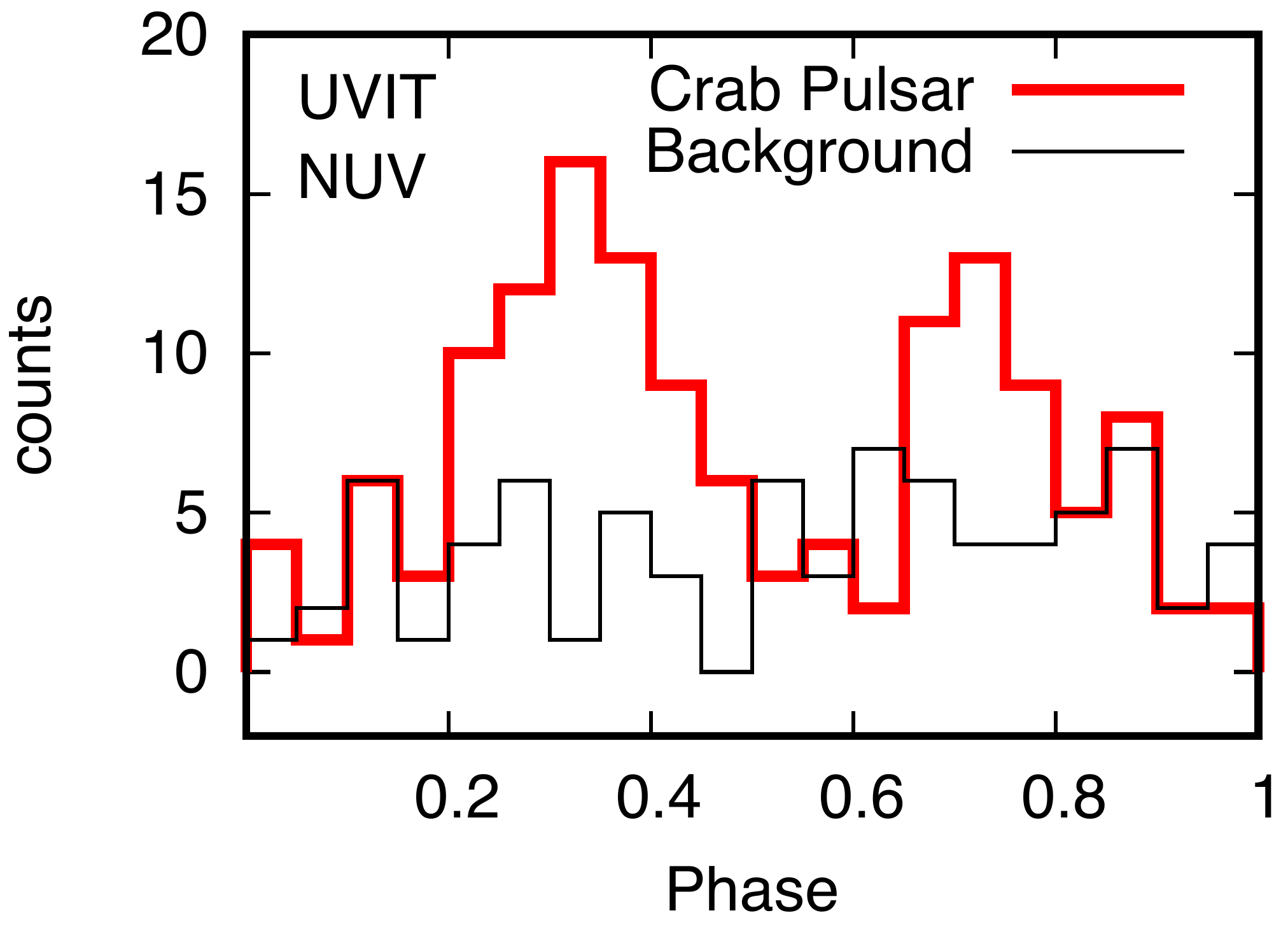}
\caption{Folded pulse profile of the Crab pulsar obtained from a 221-s observation in the NUV channel (thick red line) compared with that from background events acquired over the same time (thin black line).  Modulation due to the pulsar is clearly seen. The event lists were extracted by the UVIT Payload Operation Centre at Indian Institute of Astrophysics, Bengaluru and made available by S.N. Tandon.}
\label{fig:crabpsr_nuv}
\end{figure}

Pulsations of the Crab pulsar cannot be resolved by the frame rate of the SXT, but the small window mode of the UVIT is capable of doing so.  Figure~\ref{fig:crabpsr_nuv} shows a pulse profile obtained from an UVIT observation carried out with a 100$\times$100 window, providing a time resolution of 1.6~ms.  Data from the NUV channel was used, and a segment of length $\sim 221$~s was chosen, largely free of time jumps.  The figure compares the folded profile of events extracted from within a circle of 5~arcsec radius centred on the pulsar, with that from a similar circle excluding the pulsar position. The modulation due to the pulsar is clearly evident.  This demonstrates the potential of the UVIT for timing studies, but a method to detect and correct for clock jumps will be required to obtain longer integrations for such purposes.

\begin{figure}
\includegraphics[width=\columnwidth]{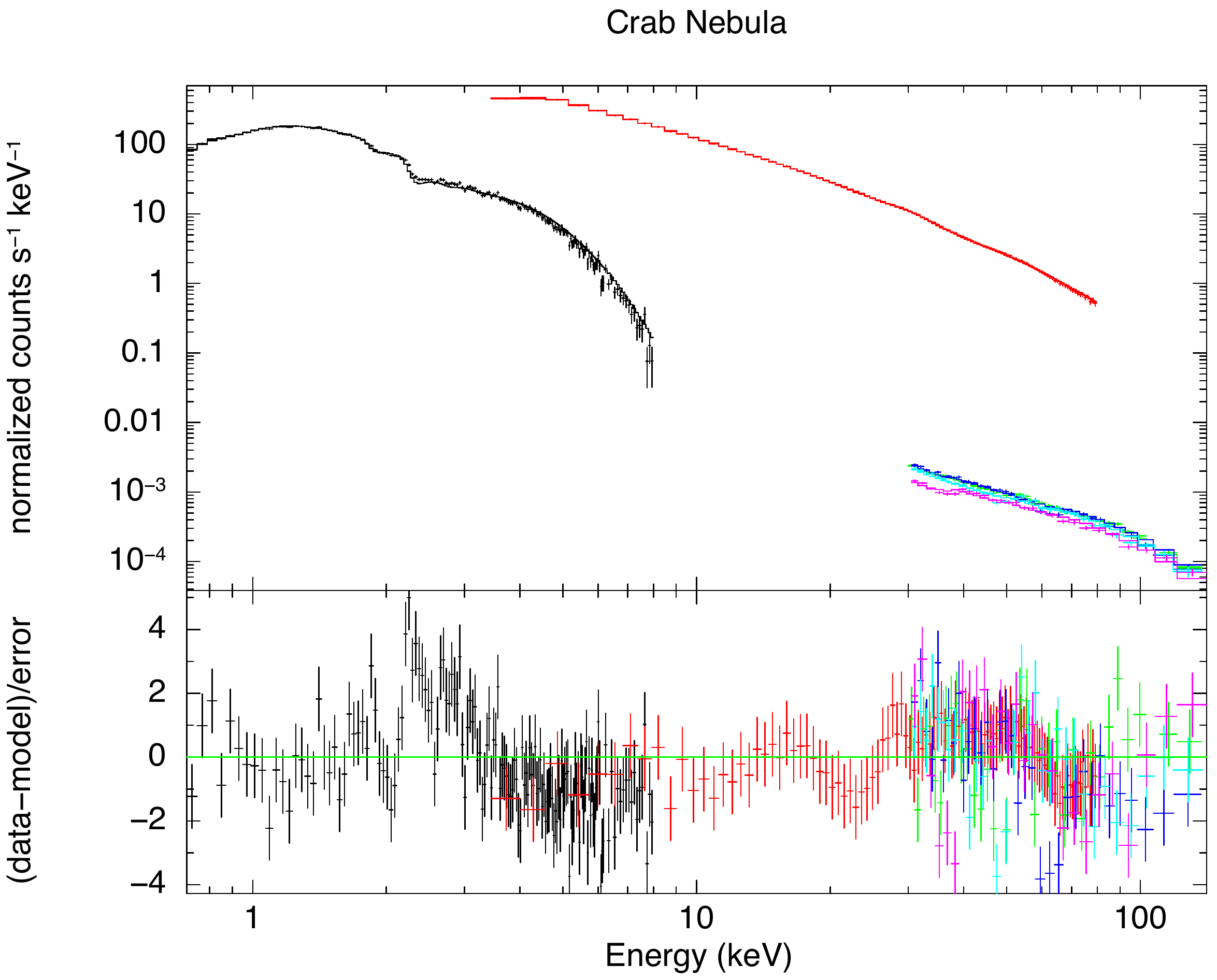}
\caption{Spectrum of the Crab nebula observed by X-ray instruments aboard AstroSat.  Count rates in the SXT (black, up to 8~keV), one unit of the LAXPC (red, 3--80~keV)  and three quadrants of the CZTI (blue, cyan and magenta, 30--110~keV) are shown.  A common spectral model of a single power law of photon index 2.11 and absorbing column $N_{\rm H} = 3.5\times 10^{21}$~cm$^{-2}$ is fitted to all the measurements using the respective response matrices.  Fit residuals shown in the lower panel demonstrate that the responses are by and large well determined. Figure courtesy Anjali Rao and K.P. Singh.}
\label{fig:crabspec}
\end{figure}

Apart from timing, another important strength of AstroSat is broadband X-ray spectroscopy.  The combination of the SXT, LAXPC and CZTI can, for bright sources, provide an excellent simultaneous coverage of the spectrum all the way from 0.3 to $\sim 100$~keV.  For example, a joint spectral fit of the Crab nebula is shown in fig~\ref{fig:crabspec}.  Count rates detected by the SXT, one unit of LAXPC and three quadrants of CZTI are displayed in this diagram along with fits with the corresponding spectral response, for a common  source spectrum model of a single power law of photon index 2.11 with an absorption equivalent to that produced by solar abundance gas with hydrogen column density $N_{\rm H} = 3.5\times 10^{21}$~cm$^{-2}$.  Residuals from the model fit are also shown, which are largely featureless.  Improvements to the response matrices are still being made to account for the small systematic deviations seen at a few places.  The net exposure in this spectrum is about 25~ks.

\begin{figure}
\includegraphics[angle=270,width=\columnwidth]{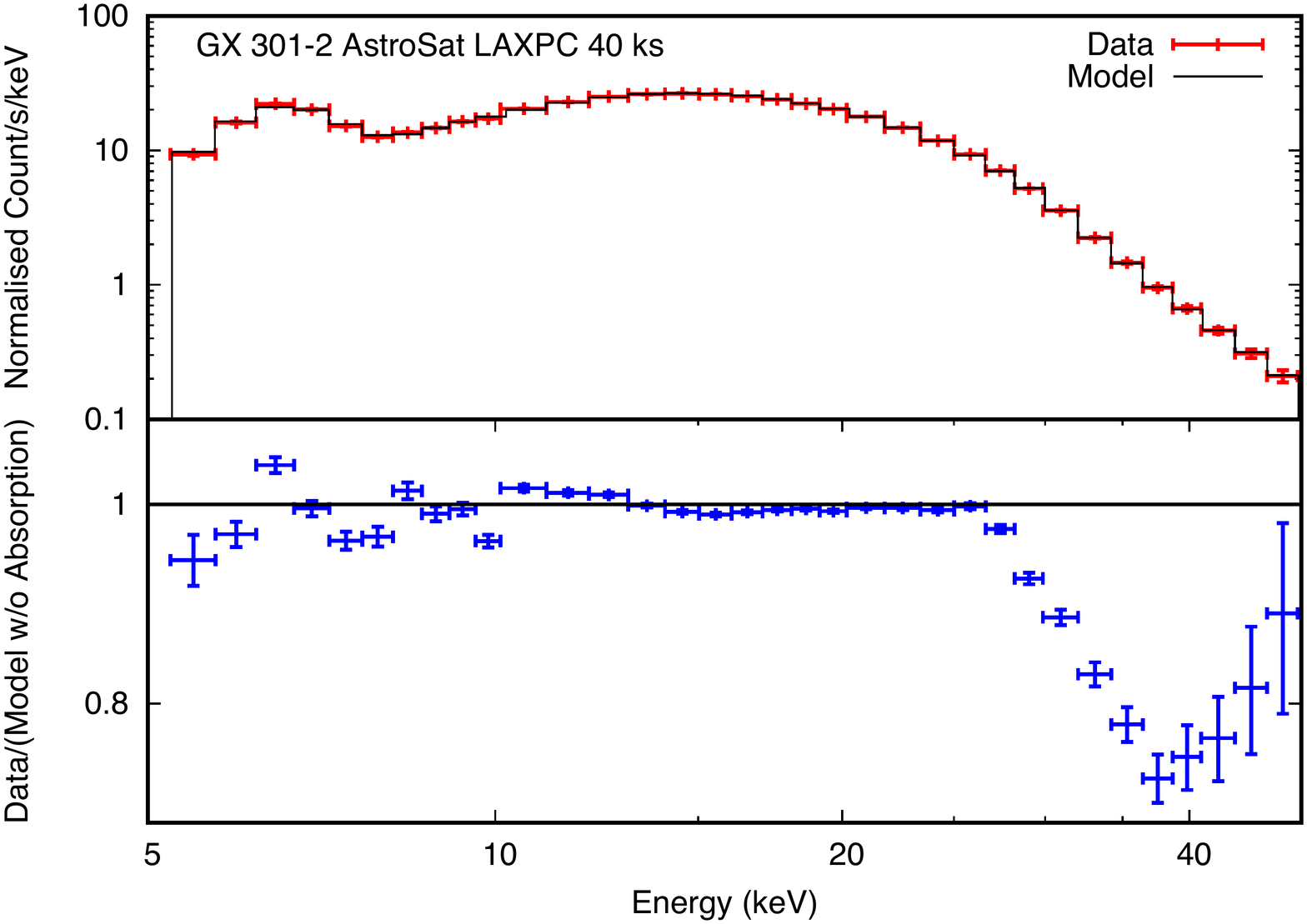}
\caption{Phase averaged spectrum of the Be X-ray binary GX~301-2 from one unit of AstroSat LAXPC. Upper panel shows the observed spectrum (points with error bars, red) and the fitted spectral model (histogram, black line).  The model consists of an absorbed continuum, an iron emission line and a cyclotron absorption feature.  The lower panel shows the residuals from the fitted model, if the absorption line is omitted from the model definition.  The cyclotron feature at $\sim 37.7$~keV is detected with high significance. Data analysis performed by Suman Bala.}
\label{fig:gx301}
\end{figure}

The final result we present here is the LAXPC spectrum of the X-ray binary GX 301-2, demonstrating the detection of a cyclotron resonance scattering feature.  This is a Be X-ray binary with the accretor being a strongly magnetised neutron star.  The neutron star has a spin period of $\sim 687$~s and is known to exhibit a cyclotron resonance scattering feature at $\sim 35$~keV, signifying a field strength of $\sim 3\times 10^{12}$~G in the emission region (Doroshenko {\em et al} \nocite{doroshenko+2010} 2010). Presented in figure~\ref{fig:gx301} is the phase-averaged spectrum of this source derived from a 40~ks observation with one of the LAXPC units.  The source spectrum is fitted with a model consisting of an absorbed power law with Fermi-Dirac cutoff, an iron line at $\sim 6$~keV and a gaussian cyclotoron absorption.  The best fit value of the line centre is $37.7\pm 2$~keV, with a width $\sigma \approx 5$~keV.  The lower panel of the figure shows the fit residuals without including the absorption component, demonstrating the statistical significance of the line feature in the spectrum.  Phase resolved spectroscopy of the cyclotron feature is being carried out from this data set, and will be reported elsewhere (S. Bala {\em et al} 2017, in preparation).  Such observations, coupled with simultaneous timing with AstroSat, are set to improve our understanding of the accretion-driven spin and magnetic evolution of neutron stars (Srinivasan and van den Heuvel \nocite{sh:1982} 1982, Radhakrishnan and Srinivasan \nocite{rs:1982} 1982, Bhattacharya and van den Heuvel \nocite{bh:1991} 1991, Srinivasan \nocite{srini:2010} 2010).

\section{Summary}
In conclusion, one can state that AstroSat has demonstrated its capability of sensitive observations for compact object science.  Event mode data recording in all bands of AstroSat lends the data to very flexible analysis that can be easily customised to the users' requirements.  The majority of the observing time of AstroSat is being devoted to diverse observations of various compact objects.  The observed data has just started to reach the proposers, and one expects a variety of results to emerge shortly.  AstroSat can clearly be counted among the most powerful instruments for compact star research till date.

A variety of material including documents, software and tools related to the usage of AstroSat is provided online at the web portal of the AstroSat Science Support Cell\footnote{http://astrosat-ssc.iucaa.in}.  Interested readers are encouraged to visit this website for further information.

\section*{Acknowledgements}
AstroSat was built and is operated by a large team of people from a number of institutions in India and abroad.  Tireless effort by numerous contributors over two decades has enabled the mission to realise its potential.  The fabrication, launch and operations have been managed and funded by the Indian Space Research Organisation (ISRO).  Science payloads were built by a consortium of institutions including the Tata Institute of Fundamental Research, Indian Institute of Astrophysics, Raman Research Institute, Inter-University Centre for Astronomy and Astrophysics, Physical Research Laboratory, University of Leicester and the Canadian Space Agency, in addition to multiple centres of ISRO.

The idea of the mission was initiated in 1996 by Dr. K. Kasturirangan, then chairman of ISRO and was developed under the stewardship of Prof. P.C. Agrawal.  The present author was  introduced to the AstroSat mission by Prof. G. Srinivasan, who was among those who contributed to the development of the initial mission concept.  He, along with Dr. George Joseph and other members of the AstroSat Payload Monitoring Committee guided and nurtured the development of the AstroSat payloads during the early years. Prof. Srinivasan was also instrumental in raising the global awareness of this mission through tireless discussions in numerous national and international fora.  

Prof. P.C. Agrawal led the mission as the Principal Investigator until passing on the responsibility to Dr. S. Seetha.  He also led the development of the LAXPC payload, which was later taken over by Prof. R.K. Manchanda and then Prof. J.S. Yadav.  The UVIT instrument was led initially by Prof. N. Kameswara Rao and later by Prof. S.N. Tandon.  Dr. S. Seetha led the Scanning Sky Monitor instrument until this role was assigned to Dr. M.C. Ramadevi.  The SXT instrument has been led by Prof. K.P. Singh and the CZT Imager by Prof. A.R. Rao.  Canadian Space Agency provided the detectors of the UVIT instrument, and this partnership has been managed by Prof. J. Hutchings.  The CCD camera of SXT was built at the University of Leicester and liaised by Dr. Gordon Stewart.

V. Koteswara Rao, followed by K. Suryanarayana Sarma were the Project Directors of AstroSat at ISRO, and were primarily responsible for the overall development and realisation of the mission.  Their contributions have been absolutely critical in making the mission a reality.  Unwavering support by successive chairpersons of ISRO, Dr. K. Kasturirangan, Dr. G. Madhavan Nair, Dr. K. Radhakrishnan and Dr. A.S. Kiran Kumar has been instrumental in ensuring the success of the mission. This paper uses data from the AstroSat mission of the Indian Space Research Organisation (ISRO), archived at the Indian Space Science Data Centre (ISSDC).

\end{document}